# Current-Voltage Characteristics of Long-Channel Nanobundle Thin-Film Transistors: A 'Bottom-up' Perspective

N. V. Pimparkar[1], S. Kumar[2], J. Y. Murthy[2] and M. A. Alam[1]
[1]School of Electrical and Computer Engineering, Purdue University, West Lafayette, IN 47907-1285, USA.
[2]School of Mechanical Engineering, Purdue University, West Lafayette, IN 47907-2088, USA.

*Abstract*— By generalizing the classical linear response theory of 'stick' percolation to nonlinear regime, we find that the drain current of a Nanobundle Thin Film Transistor (NB-TFT) is described under a rather general set of conditions by a universal scaling formula $I_D = A/L_S\ \xi(L_S/L_C, \rho_S L_S^2) \times f(V_G, V_D)$, where $A$ is a technology-specific constant, $\xi$ is function of geometrical factors like stick length ($L_S$), channel length ($L_C$), and stick density ($\rho_S$) and $f$ is a function of drain ($V_D$) and gate ($V_G$) biasing conditions. This scaling formula implies that the measurement of full I-V characteristics of a *single* NB-TFT is sufficient to predict the performance characteristics of any other transistor with arbitrary geometrical parameters and biasing conditions.

*Index Terms*— Carbon Nanotube, Network Transistor, Inhomogeneous Percolation Theory, Thin Film Transistor.

## I. INTRODUCTION

Over last several years many research groups have developed NB-TFTs based on percolating network of randomly-oriented, finite-length Silicon nanowires (NW) and Carbon nanotubes (NT), as shown in Fig. 1. Potential applications include macroelectronics systems such as displays, e-paper, e-clothing, biological and chemical sensors, conformal radar, solar cells and others [1-8]. These applications often require higher performance than a-Si and Organics and lower temperature processing than single crystal Si and p-Si for flexible substrates such as plastics. Therefore NB-TFTs based on NW/NTs are expected to be better suited for these high performance macroelectronic applications.

The carrier transport characteristics in NB-TFTs (Fig. 1) has been previously modeled and compared to experiments in linear response regime ($V_{TH} < V_G < V_{DD}$, $V_D \sim small$) [9]. Instead of using a traditional 'top-down' phenomenological effective mobility model to describe carrier transport[10], Ref. [9] uses a 'bottom-up' approach where the properties of the thin film reflects the percolating geometry of NW/NT network. Specifically, in the linear response regime, the charge density induced in each NW/NT, $n = C_{OX}(V_G - V_{TH})$, is a constant independent of $V_D$. Therefore, the transport properties of NB-TFT in the linear-response regime are readily described by the theory of 2D homogenous (constant conductivity) stick percolating networks.

The performance limit of NB-TFTs, however, is dictated by the transistor characteristics at high bias (nonlinear) regime ($V_{TH} < V_G$, $V_D < V_{DD}$). High interface trap density ($N_{IT}$) [11] and large hysteresis associated with current-generation of NB-TFTs make stable measurement of their high-bias characteristics difficult. Therefore, predictive simulation of high-bias operation can be used to establish both the performance limits of NB-TFTs as well as relative importance of various device parameters. The fundamental issue is this: although the spatial geometry of the NW/NT network does not change with bias, the 'low-bias' assumption of constant conductivity and local homogeneity along the channel is no longer valid at 'high-bias' regime (see Fig. 1b vs. Fig. 1c). Since the traditional percolation theory demand global spatial homogeneity, the classical theory can not be used to analyze the high bias regime for NB-TFTs.

In this paper, we generalize the classical percolation theory to nonlinear (high-bias) regime and establish the performance limits for NB-TFTs using self consistent solution of Drift-Diffusion and Poisson equations. Surprisingly we find that the conductance exponent $m$ ($I_D \sim 1/L_C^m$) in high-bias inhomogeneous case is *exactly* the same as those in low bias homogeneous condition [9]. This universal scaling theory implies that the measurement of full I-V characteristics of a single transistor can be used to predict the high-bias transistor characteristics of any other NB-TFTs with arbitrary geometrical parameters (e.g., channel length, stick length, etc.) and operating conditions. Previously, we have established the theoretical basis of the scaling formula for short-channel NB-TFTs with $L_C < L_S$ in [12]. In the next section we show numerically that the scaling formulation holds even for technologically-relevant long-channel NB-TFTs with $L_C > L_S$.

## II. COMPUTATIONAL MODEL

Fig. 1a shows a typical NB-TFTs assembled with a bundle of NW/NT of length $L_S$, isotropically oriented $(0 \leq \theta < 2\pi)$ onto the surface of gate oxide. The probability of germination of NW/NT at each location is dictated by the average density of tubes, $\rho_s$. For NBTs with $L_C > L_S$, the stick-stick interaction is important and analytical solution is not possible. The 'low bias' assumptions [9] that charge density, $n$, is constant along the channel and is independent of $V_D$ do not apply at high-bias conditions. Therefore, $n(V_D, V_G)$ must be determined self-consistently by solving the drift-diffusion equations (appropriate for $L_C > 1\ \mu m$) and the Poisson equation. In the 'bottom-up' description of the channel, the DD-Poisson equations are generalized for NB-TFTs as:

$$\left.\begin{array}{l}\dfrac{d^2\Phi}{ds^2}+\dfrac{\rho}{\varepsilon}=0 \\ \nabla \cdot J_p = 0 \\ \nabla \cdot J_n = 0\end{array}\right\} \rightarrow \begin{array}{l}\sum_{i=1}^{N}\left(\dfrac{d^2\Phi_i}{ds^2}+\dfrac{\rho_i}{\varepsilon}-\dfrac{(\Phi_i-V_G)}{\lambda^2}+\sum_{j\neq i}\dfrac{(\Phi_j-\Phi_i)}{\lambda_{ij}^2}\right)=0, \\ \sum_i\left(\nabla \cdot J_{pi}+\sum_{j\neq i}C_{ij}^p(p_j-p_i)\right)=0, \\ \sum_i\left(\nabla \cdot J_{ni}+\sum_{j\neq i}C_{ij}^n(n_j-n_i)\right)=0,\end{array} \quad (1)$$

where, $N$ is total number of NT/NW, $s$ is in the direction of individual NW/NT, $\rho$ is total charge density, and the term $-(\Phi - V_G)/\lambda^2$ (the well known parabolic approximation [13, 14]) introduces the effect of back gate, where $\lambda$ is effective screening length with $\lambda^2 = \varepsilon_{CNT} T_{ox} d / \varepsilon_{OX}$. For typical transistor parameters of $d \sim 2\ nm$ is the thickness of the nanobundle film, $T_{ox} \sim 250\ nm$ is thickness of gate oxide, $\varepsilon_{CNT} \sim 5$ [15] and $\varepsilon_{OX} \sim 3.9$ are dielectric constants for CNT network and gate oxide, respectively, gives $\lambda \sim 44\ nm$. The parabolic approximation is valid in this case because the condition that $L_C >> \lambda >> d$ is satisfied. The term $(\Phi_j - \Phi_i)/\lambda_{ij}^2$ is stick-stick interaction with screening length $\lambda_{ij}$ where a node on stick $i$ intersects a node on stick $j$. The intersecting nodes act as tiny gates for each other modifying the potential and carrier concentrations [16]. Further, transport is essentially 1D (along the tube) with the additional term $C_{ij}^n(n_j - n_i)$ in the continuity equation representing charge transfer between nanosticks at the point of intersection. Here higher value of $C_{ij}^{n,p} = G_0/G_1$ implies better electrical contact, where $G_0$ and $G_1$ is mutual and self conductances of the tubes [9].

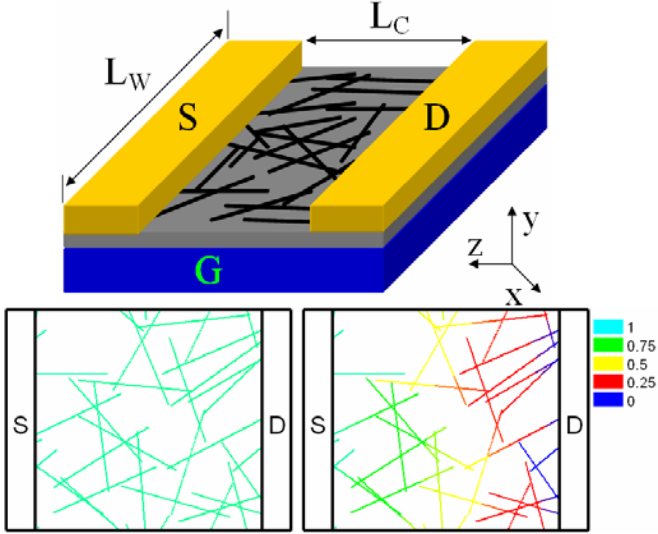

Fig. 1. (a) A schematic of NBT: Geometry of nanobundle network transistor. Distribution of normalized conductance along the sticks in the random network in (b) low and (c) high bias, respectively. The conductance is uniform along the channel in low bias while it varies nonlinearly in the high bias condition resulting in an inhomogeneous percolating network and the contours of conductance are not straight lines parallel to S/D as in conventional MOSFET.

## III. RESULTS AND DISCUSSION

It was shown in [12] that in the 'short-channel' limit of $L_C < L_S$ and at low stick density, the NW/NTs behave as individual transistors connected in parallel and therefore the ratio of the $I_D$ for any two bias points is independent of the geometry of the NB-TFTs. This implies the scaling relationship that

$$I_D = \frac{A}{L_S}\xi\left(\frac{L_S}{L_C}, \rho_s L_S^2\right) \times f(V_G, V_D), \quad (2)$$

where the proportionally constant $A$ depends on oxide capacitance $C_{ox}$, tube diameter $d$ [17], and stick-stick interaction parameter, $C_{ij}^{n,p}$. And $\xi$ and $f$ are functions of geometrical parameters ($L_S$, $L_C$, $\rho_S$) and bias conditions ($V_D$, $V_G$), respectively. Fig. 10 in Ref. [12] shows that this factorization is consistent with experimental data.

For long-channel NB-TFTs with $L_C > L_S$ (see Fig. 1a), the individual sticks can not bridge the channel by themselves and stick-stick interaction becomes important ($C_{ij}^{n,p} \neq 0$). Fig. 2 summarizes the self-consistent solution of (1) for different bias conditions (Fig. 2a) and various geometrical parameters (Fig. 2b, c) for long-channel NB-TFTs. Each point of Fig. (2) reflects the average solution of ~200 statistical samples and generation of the figure requires approximately 15 hrs on a 200 nodes of a cluster with 3.2 GHz, 64-bit Intel Irwindale processors with 4GB of memory. Remarkably the results in Fig. 2 indicate that the scaling formula (2) holds for arbitrary geometrical and biasing conditions even *in the long-channel limit of $C_{ij}^{n,p} \neq 0$*. Specifically, (2) requires that for the geometrical scaling factor, $\xi(L_S/L_C, \rho_S L_S^2)$ remain invariant (up to a scaling factor) of biasing conditions which is easily confirmed by comparing Fig. 2b and 2c. Indeed, we find that

$$\xi\left(\frac{L_S}{L_C}, \rho_s L_S^2\right) = \left(\frac{L_S}{L_C}\right)^{m(\rho_s L_S^2)} \quad (3)$$

where $m$ is a universal exponent of stick percolating system. For densities much higher than percolation threshold ($\rho_S L_S^2 >> \rho_{th} L_S^2 = 4.236^2/\pi$ [18]), the network behaves as a 2D conductor with $m = 1$. Together with Eq. (4) below we find that Eq.(2) reduces to the classic 'square law' as expected. But for densities near percolation threshold ($\rho_S L_S^2 \sim 4.236^2/\pi$) the exponent takes the value $m \sim 1.8$ (Fig. 2b). Moreover, Fig. 2a shows that the bias-dependent scaling function

$$f(V_G, V_D) = \left[(V_G - V_{TH})V_D - \beta V_D^2\right] \quad (4)$$

is independent of geometrical parameters ($\beta \sim 0.5$), again satisfying Eq. (2). It is hardly surprising that the voltage scaling function $f(V_G, V_D)$ would follow the classical 'square-law' formula at very low ($\rho << \rho_{th}$) and at very high densities ($\rho >> \rho_{th}$). After all, for ($\rho << \rho_{th}$) the stick-stick interaction is negligible, therefore the system behaves as an independent collection of 1D (long-channel) conductors and the conclusions of Ref. [12] apply. At $\rho >> \rho_{th}$ the percolating network approximates a classical 2D homogenous thin film and once again the classical MOSFET formula should hold. *The real surprise and the most significant finding of our analysis is that Eq. (2) holds for arbitrary stick density above and below the percolation threshold.* Our analysis indicates (details to be published elsewhere) that this robust density independent voltage scaling arises from the fact that, given the 1D approximation to Poisson equation in (1), any random network of sticks can be reduced to a single (homogeneous) thin-film strip of effective width *W*. The scaling holds because the geometrical factor $\xi$ measures this effective width of the transistor at various tube densities, stick lengths, and channel lengths.

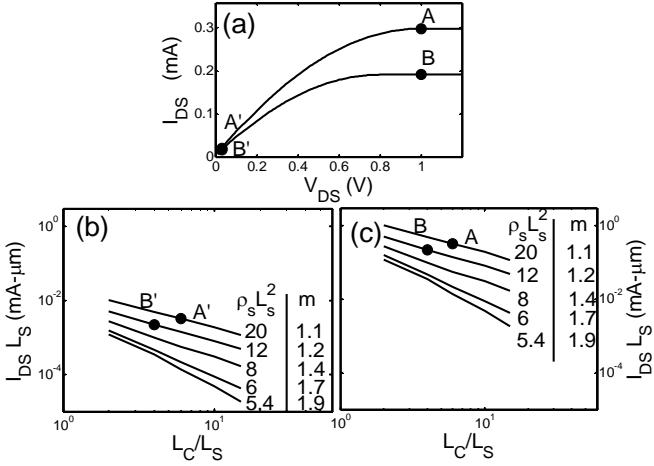

Fig. 2. (a) $I_{DS}$-$V_{DS}$ for 2 different random networks *A* and *B*, Drain current vs. channel length for network transistors with channel length, $L_C$ > stick length, $L_S$ for (b) low and (c) high bias, respectively, in strong coupling limits. The current exponents, *m*, for high bias regimes are *exactly* same as low bias regime, and Eq. (2) requires that the ratio of currents at two different biasing conditions is independent of geometrical parameters for any particular percolating random network, i.e. $I_A/I_{A'} = I_B/I_{B'}$.

Our results imply that once $V_{TH}$ and $\beta$ is determined (for Eq. 4) by $I_D$-$V_D$ and $I_D$-$V_G$ measurement and *m* is determined from Fig. (2b) for particular $L_S$, $L_C$, and $\rho_S$, one can readily determine the technology specific constant *A*. Given *A*, $V_{TH}$ and $\beta$, we can determine (by Eq. 2) the transistor performance of any other transistor of arbitrary $L_S$, $L_C$, and $\rho_S$, (Fig. 2b) and biasing condition. This scaling formula could therefore reduce the technology development and characterization time significantly. Second, Eq. (2) provides a *'bottom-up'* definition of effective-mobility, $\mu_{eff} \sim (dI_D/dV_G/V_D)(L_S/\xi)/(L_W C_{OX})$, the value of which is independent of $L_C$ and can be used to compare experimental data from various laboratories. For very high density networks, $\mu_{eff}$ reduces to conventional mobility equation as $m = 1$ and $\xi = L_S/L_C$ in Eq. (3). Finally, Eq. (1) and (2) can be used to compute $f_{max} = I / CV$ to establish ultimate performance limits of NB-TFTs free from nonideal factors like hysteresis or interface traps which gives ~ 1GHz of device speed for NB-TFT in [3] for $L_C = 1$ $\mu m$.

## IV. CONCLUSIONS:

We have generalized the linear stick percolation theory to nonlinear regime to find a scaling formula (2) to compute $I_D$-$V_D$ characteristics of NB-TFT that, once calibrated, can be used to establish performance limits of NB-TFTs of arbitrary geometry and operating conditions. Our analysis therefore would help organize experimental data from various research groups and could have significant impact on the development of NB-TFT technology.

## V. ACKNOWLEDGEMENTS:

This work was supported by the Network of Computational Nanotechnology and the Lilly Foundation.